\documentstyle [12pt]{article}
\begin{document}
\normalsize
\centerline{\bf On calculation of effective conductivity of inhomogeneous 
metals}
\vskip 5mm
\centerline{\it Inna M. Kaganova}
\small
\centerline{Institute for High Pressure Physics Russian Academy of 
Sciences}
\centerline{142190 Troitsk, Moscow Region; e-mail:
kaganova@hppi.troitsk.ru}
\normalsize
\begin{abstract}
In the framework of the perturbation theory an expression suitable 
for calculation of the effective conductivity of 3-D inhomogeneous 
metals is derived. Formally, the final expression is an exact 
result, however, a function written as a perturbation series 
enters the answer. More accurately, when statistical properties of 
the given inhomogeneous medium are known, our result provides the 
regular algorithm for calculation of the effective conductivity up 
to an arbitrary term of the perturbation series. As examples, we 
examine (i) an isotropic metal whose local conductivity is a 
Gaussianly distributed random function, (ii) the effective 
conductivity of polycrystalline metals.
\end{abstract} 



\section{Introduction}

Rather often macroscopic properties of stochastically 
inhomogeneous media, which are isotropic in average, can be 
described in the framework of different models of effective 
isotropic medium. Calculation of the effective conductivity (EC) 
is one of the well-known problems of theory of conductive media. 
It has been studied theoretically for many years (see, e.g., the 
review paper by A.G.Fokin \cite{1}). However, in the general case 
up to now this problem has not been solved yet. 

Exact solutions for effective characteristics of stochastically 
inhomogeneous media can be found very rarely. One of such examples 
is the calculation of EC of some two-dimensional inhomogeneous 
media \cite{2}. The existence of this result is due to a specific 
symmetry transformation allowed by the equations of the problem. 
Recently in the frequency region of the local impedance (the 
Leontovich) boundary conditions applicability the exact solution 
for the effective surface impedance of inhomogeneous metals was 
obtained \cite{3,4}. Sometimes effective characteristics can be 
estimated qualitatively. For example, there are no regular methods 
allowing us to calculate the effective conductivity of three-
dimensional (3-D) polycrystals accurately, but some geometrical 
reasoning suggested in Ref.\cite{5}, allowed the authors to 
estimate EC of strongly anisotropic polycrystals.

When the exact solution for an effective value ${\psi}_{ef}$ of a 
characteristic $\psi$ cannot be found, perturbation theory 
calculations may be useful. In a stochastically inhomogeneous 
medium $\psi$ is a stochastic function of position $\bf r$. A 
natural zero order approximation for ${\psi}_{ef}$ is $<\psi>$ 
that is of $\psi ({\bf r})$ averaged over all possible 
realizations of the medium. However, when calculating $<\psi>$, we 
do not take into account corrections due to the spatial 
fluctuations of $\psi ({\bf r})$. 

From our point of view, the most accurate and physically 
meaningful method to take account of spatial fluctuations goes 
back to the work of I.M.Lifshitz and L.N.Rosenzweig \cite{6}. They 
proposed to start from differential equations of the problem 
coefficients of which are random functions of position. Averaging 
these equations we derive equations for the averaged fields, 
which allow us to determine the effective characteristic. Usually 
this approach works when the inhomogeneity is small and 
perturbation theory is applicable. Previously it was used to 
calculate a lot of static and dynamic characteristics of 
inhomogeneous media. As an example, we cite Refs. 
\cite{7,8,9,10,11}. In all these works only the first nonvanishing 
term taking account of spatial fluctuations was calculated. For a 
characteristic $\psi$ this term is of the order of $<{(\psi -
<\psi>)}^2>$. Usually, it does not depend on the 
correlations of the values of $\psi$ in different points of the 
medium. 

It is much more difficult to calculate the corrections of the 
higher orders in $(\psi ({\bf r}) - <\psi>)$. The calculation 
involves step by step derivation of equations for each correction 
of the order $n>2$. In addition, the high order corrections depend 
on the statistical properties of the medium.

In this paper with the aid of the perturbation theory we derive an 
expression suitable for calculation of EC ${\sigma}_{ef}$ of 3-D 
inhomogeneous metals. The basic formula for ${\sigma}_{ef}$ is 
obtained in Section 2. Formally, in the framework of the 
perturbation theory applicability, the final expression, Eq.(15), 
is exact. However, a function written as a perturbation series 
enters the answer. So, more accurately, when statistical 
properties of the given inhomogeneous medium are known, our result 
provides the regular algorithm for calculation of EC up to an 
arbitrary term of the perturbation series. We also write down the 
perturbation theory expression for EC of two-dimensional (2-D) 
conducting media. As an example, we examine an isotropic metal 
whose local conductivity is a Gaussianly distributed random 
function, and calculate the forth order correction to the 
effective conductivity. 

In Section 3 we consider polycrystalline metals. To verify our 
perturbation theory, in Subsection 3.1 we compare the perturbation 
theory calculation of ${\sigma}_{ef}$ for 2-D polycrystals with 
the exact result of \cite{2}. In Subsection 3.2 EC of 3-D 
polycrystals is discussed. We calculate EC up to the third order 
term in the parameters of anisotropy and outline the region of the 
perturbation theory applicability. Also we examine the dependence 
of the third order term on the statistical properties of the 
polycrystal.
 
\section{Perturbation Theory for Effective Conductivity of 3-D 
Inhomogeneous Metals}

We consider a stochastically inhomogeneous metal that is isotropic 
and homogeneous in average. The elements of the local conductivity 
(LC) tensor ${\sigma}_{ik}$ are stochastic functions of position 
${\bf r}$. By $<...>$ denote the ensemble average over all 
possible realizations of the medium. The averaged conductivity 
$<{\sigma}_{ik}({\bf r})> = <\sigma >{\delta}_{ik}$, and 
$$
{\sigma}_{ik}({\bf r}) = <\sigma >({\delta}_{ik} + 
{\Delta}_{ik}({\bf r})), \quad <{\Delta}_{ik}({\bf r})> = 0.
\eqno (1)
$$
The stochastic tensor ${\Delta}_{ik}({\bf r})$ describes the 
spatial fluctuations of LC. 

By definition, the effective conductivity ${\sigma}_{ef}$ is 
specified by equation:
$$
<{\bf j}({\bf r})> = {\sigma}_{ef}<{\bf E}({\bf r})>, \eqno (2)
$$ 
where $<{\bf j}>$ is the macroscopic direct current density and 
$<{\bf E}>$ is the uniform macroscopic electric field; 
${\bf j}({\bf r})$ and ${\bf E}({\bf r})$ are the local current 
density and the local electric field, respectively.

To calculate ${\sigma}_{ef}$, we write the electrostatics 
equations 
$$
{\rm div}{\bf j} = 0, \quad {\rm rot}{\bf E} = 0, \eqno (3)
$$
and the material equation that is the Ohm law. According to Eq.(1) 
$$
{j}_{i}({\bf r}) = <\sigma >({\delta}_{ik}+{\Delta}_{ik}
({\bf r})){E}_{k}({\bf r}). \eqno (4)
$$
Because of the tensor ${\Delta}_{ik}({\bf r})$, Eq.(3) constitute 
a system of stochastic equations for the local field 
${E}_{i}({\bf r})$.

We write ${\bf j}({\bf r})$ and ${\bf E}({\bf r})$ as ${\bf j} 
= <{\bf j}> + \delta {\bf j}$ and ${\bf E} = <{\bf E}> + \delta 
{\bf E}$ ($<\delta {\bf j}>=<\delta {\bf E}>=0$); $\delta {\bf j}$ 
and $\delta {\bf E}$ are position dependent stochastic current and 
stochastic field, respectively. Substituting these 
expressions in Eq.(4) we obtain after averaging
$$
<j_i> = <\sigma >(<E_i> + J_i), \quad 
J_{i} =  <{\Delta}_{ik}({\bf r}){\delta E}_k({\bf r})>. \eqno (5)
$$
Comparing Eq.(2) and Eq.(5), we see that the uniform vector $\bf 
J$ defines the contribution of the spatial fluctuations of LC to 
the value of EC. 

Subtracting Eq.(5) from Eq.(4) we have
$$
{\delta j}_i = <\sigma >({\delta E}_{i} + {\Delta}_{ik}<E_k> + 
D_i), \quad D_i = {\Delta}_{ik}{\delta E}_{k} - J_i. \eqno (6)
$$
The components of the vector $\bf D$ at least quadratic in powers 
of the elements of the tensor ${\Delta}_{ik}$. When calculating EC 
up to the first nonvanishing term taking account of spatial 
fluctuations, the vector $\bf D$ in Eq.(6) has to be omitted. 
However, just this vector defines corrections of higher 
orders.

To calculate the vector $\bf J$ we use the perturbation theory. We 
present the stochastic field ${\delta E}_{i}({\bf r})$ as a series 
in powers of the elements of the stochastic tensor 
${\Delta}_{ik}({\bf r})$: 
$$
{\delta E}_{i}({\bf r}) = \sum_{n=1}^{\infty}
{\delta E}_{i}^{(n)}({\bf r}).
\eqno(7.a) 
$$
and seek the vector $J_i$ as the series 
$$
J_i = \sum_{n=2}^{\infty}J_i^{(n)},  \quad 
J_i^{(n)} = <{\Delta}_{jk}({\bf r}){\delta E}_{i}^{(n-1)}
({\bf r})>. \eqno (7.b)
$$

With regard to Eq.(6) from the electrostatic equations (3) it 
follows that the Fourier coefficient of the stochastic field 
${\delta E}_i({\bf r})$ is ${\delta E}_{i}({\bf k}) = -
{\kappa}_{i}{\kappa}_{j}({\Delta}_{jk}({\bf k})<E_k>+
D_j({\bf k}))$, where ${\kappa}_{i}=k_i/k$. Consequently, the 
Fourier coefficients of ${\delta E}_{i}^{(n)}({\bf r})$ are
$$
{\delta E}_{i}^{(1)}({\bf k}) = -{\kappa}_{i}{\kappa}_{j}
{\Delta}_{jk}({\bf k})<E_k> \quad {\rm and}\quad
{\delta E}_{i}^{(n)}({\bf k}){|}_{n>1} = -{\kappa}_{i}{\kappa}_{j}
D_j^{(n)}({\bf k}), \eqno (8)
$$
where $D_j^{(n)}({\bf k})$ is the Fourier coefficient of 
$D_j^{(n)}({\bf r}) = {\Delta}_{jk}({\bf r})
{\delta E}_{k}^{(n-1)}({\bf r}) - J_j^{(n)}$. 

Equations (7) and (8) allow us to write the n-th term of the 
series (7.b) as a sum of the lower orders terms $J_i^{(m)}$ ($2 
\le m < n$). Indeed, for $n \ge 2$ from Eq.(8) it follows that
$$
{\delta E}_{i}^{(n)}({\bf r}) = -\left\{\int\int {\rm d}^{3}k_{n}
{\rm d}^{3}k_{n-1}{q}_{ij_n}^{(n)}{\Delta}_{j_nl_{n-1}}
({\bf k}_{n}-{\bf k}_{n-1}){\delta E}_{l_{n-1}}^{(n-1)}
({\bf k}_{n-1})
{\rm e}^{i{\bf k}_n{\bf r}} - \frac{1}{3}{J}_{i}^{(n)}\right\},
\eqno (9)
$$
where $q_{ik}^{(n)} = {\kappa}_{i}^{(n)}{\kappa}_{k}^{(n)}$. 

Next, we use Eq.(9) to write down the expression for 
${\delta E}_{i}^{(n-1)}({\bf r})$. We Fourier 
analyze ${\delta E}_{i}^{(n-1)}({\bf r})$ and substitute the 
result in Eq.(9). Then the term proprtional to $J_i^{(n-1)}$ 
appears in the right-hand side of Eq.(9) and the Fourier 
coefficient ${\delta E}_{l}^{(n-2)}({\bf k})$ enters the integral 
term of this equation.

The next steps are obvious. In successive order we decrease the 
superscripts in the expressions for the stochastic fields 
$\delta{\bf E}^{(m)}$ entering the integral term of Eq.(9). At the 
last step the coefficient ${\delta {\bf E}}^{(1)}({\bf k}_{1})$ 
defined by Eq.(8) enters the equation. 

If we use the expression for ${\delta E}_{i}^{(n)}({\bf r})$ 
obtained as the result of the aforementioned decreasing procedure 
when calculating $J_i^{(n+1)}$ and replace the superscript $(n+1)$ 
by $m$, we have
$$
\sum_{l=2}^{m} w_{ik}^{(m-l)}J_{k}^{(l)} = -3w_{ik}^{(m)}<E_k>.
\eqno (10)
$$
In Eq.(10) $w_{ik}^{(0)} = {\delta}_{ik}$, $w_{ik}^{(1)} = 
- <{\Delta}_{ik}({\bf r})> /3= 0$ and for $n \ge 2$ 
$$
w_{ik}^{(n)} = {(-1)}^{n}\frac{1}{3{(2\pi )}^{3(n-1)}}\int ...\int 
{\rm d}^{3}r_{1}...{\rm d}^{3}r_{n-1}
\int ...\int {\rm d}^{3}k_{1}...{\rm d}^{3}k_{n-1}
q_{l_1j_1}^{(1)}...q_{l_{n-1}j_{n-1}}^{(n-1)} \times
$$
$$ 
<{\Delta}_{il_1}({\bf r})
{\Delta}_{j_1l_2}({\bf r}_1)...{\Delta}_{j_{n-1}k}({\bf r}_{n-1})>
{\rm e}^{i{\bf k}_1({\bf r}-{\bf r}_1)}
{\rm e}^{i{\bf k}_2({\bf r}_1-{\bf r}_{2})}...  
{\rm e}^{i{\bf k}_{n-1}({\bf r}_{n-2}-{\bf r}_{n-1})}\eqno (11)
$$

When the medium is isotropic in average, the moments 
$<{\Delta}_{il_1}({\bf r}){\Delta}_{j_1l_2}({\bf r}_1)... 
{\Delta}_{j_2k}({\bf r}_n)>$ depend only on the modulus of the 
differences between the vectors ${\bf r}_{i}$ entering the given 
average. Then the elements of the tensors ${w}_{ik}^{(n)}$ do not 
depend on position. Since all isotropic uniform second rank 
tensors reduce themselves to the unit tensor ${\delta}_{ik}$, the 
only possible form of all the tensors $w_{ik}^{(n)}$ ($n \ge 2$) 
is
$$
w_{ik}^{(n)} = w^{(n)}{\delta}_{ik};\quad w^{(n)} = \frac{1}{3} 
w_{ii}^{(n)}, 
\eqno (12)
$$
where $ w_{ii}^{(n)}$ is the trace of the matrix $w_{ik}^{(n)}$. 
We also have $w^{(0)}=1$ and $w^{(1)}=0$.

Taking account of Eq.(12) we write down Eqs.(10) for all 
$2 \le m < \infty$:
$$
J_i^{(2)} = -3w^{(2)}<E_i>, \quad {\rm if}\; m =2; \eqno (13.a)
$$
$$
w^{(1)}J_i^{(2)} + J_i^{(3)} = -3w^{(3)}<E_i>, \quad {\rm if}\; m 
=3; \eqno (13.b)
$$
$$
w^{(2)}J_i^{(2)} + w^{(1)}J_i^{(3)} + J_i^{(4)} = -3w^{(4)}<E_i>, 
\quad {\rm if}\; m =4; \eqno (13.c)
$$
$$
w^{(3)}J_i^{(2)} + w^{(2)}J_i^{(3)} + w^{(1)}J_i^{(4)}+ J_i^{(5)} 
= -3w^{(5)}<E_i>, \quad {\rm if}\; m =5 \eqno (13.d)
$$
and so on. We add together all these equations. Then according to 
Eq.(7.b) we obtain 
$$
T_3J_i= -3(T_3-1)<E_i>, \quad T_3 = \sum_{m=0}^{\infty} w^{(m)}.  
\eqno (14)
$$
Consequently, $J_i= 3(1/T_3 - 1)<E_i>$. Finally, with regard to 
Eq.(2) and Eq.(5) we have
$$
{\sigma}_{ef} = <\sigma >(3/T_3 - 2). \eqno (15)
$$

Together with Eq.(15) we would like to present the expression for 
EC of 2-D inhomogeneous metals. Repeating the previous 
calculations with regard to the dimensionality of the problem, we 
obtain 
$$
{\sigma}_{ef} = \overline\sigma (2/T_2 - 1), \quad 
T_2 = \sum_{m=0}^{\infty} v^{(m)}
\eqno (16.a)
$$
where again $v^{(0)}=1$, $v^{(1)}=0$ and for $n \ge 2$ we have
$$
v^{(n)} = {(-1)}^{n}\frac{1}{4{(2\pi )}^{2(n-1)}}\int ...\int 
{\rm d}^{2}r_{1}...{\rm d}^{2}r_{n-1}
\int ...\int {\rm d}^{2}k_{1}...{\rm d}^{2}k_{n-1}
q_{l_1j_1}^{(1)}...q_{l_{n-1}j_{n-1}}^{(n-1)} \times
$$
$$ 
<{\Delta}_{il_1}({\bf r})
{\Delta}_{j_1l_2}({\bf r}_1)...{\Delta}_{j_{n-1}i}({\bf r}_{n-1})>
{\rm e}^{i{\bf k}_1({\bf r}-{\bf r}_1)}
{\rm e}^{i{\bf k}_2({\bf r}_1-{\bf r}_{2})}...  
{\rm e}^{i{\bf k}_{n-1}({\bf r}_{n-2}-{\bf r}_{n-1})},\eqno (16.b)
$$
${\bf r}_n$ and ${\bf k}_n$ are 2-D position and wave vectors, 
respectively. 

Equations (15) and (16) are the basic formulas we use when 
calculating EC within the framework of the perturbation theory. 
These formulas look rather simple. However, usually even for 
rather simple models of inhomogeneous conducting media we can 
calculate $T_3$ (or $T_2$) up to the desirable order in powers of 
the parameters of inhomogeneity only. 

It can appear that no significant simplification has been done. 
However, when the perturbation theory is applicable, Eq.(15) and 
Eq.(16) provide a regular method for calculation of the high order 
terms in the expression for EC.

In the end of this Section as an example, we examine an isotropic 
metal with Gaussian inhomogeneities. This example is rather 
simple, since, first, for isotropic metals the expressions for 
$w^{(n)}$ are simplified, and, second, the form of n-point 
averages entering these expressions are known. 

We suppose that LC is ${\sigma}_{ik}({\bf r}) = <\sigma >(1 + 
{\Delta}({\bf r})){\delta}_{ik}$, where $\Delta ({\bf r})$ is a 
strictly stationary, zero-mean Gaussian process \cite{12}. Then 
the two-point average is
$$
<\Delta ({\bf r})\Delta ({\bf r}_1)> = {\Delta}^{2}W_G(|{\bf r}- 
{\bf r}_{1}|),  \eqno (17.a)
$$
where ${\Delta}^{2}=<{\Delta}^2({\bf r})>$, and the Gaussian 
correlation function $W_G(r)$ and its Fourier coefficient $W_G(k)$ 
are
$$
W_G(r) = {\rm e}^{-r^2/a^2} \;{\rm and}\; W_G(k) = 
\frac{a^3}{{(2\sqrt{\pi})}^3}{\rm e}^{-{(ka)}^{2}/4}, \eqno (17.b)
$$
$a$ is the correlation radius. Next, the average of a product of 
an odd number of $\Delta ({\bf r})$ vanishes, and the average of 
an even number of $\Delta ({\bf r})$ is given by the sum of the 
products of the averages of pairs of the $\Delta ({\bf r})$'s 
taken in all possible ways, irrespective of order, e.g.
$$
<\Delta ({\bf r})\Delta ({\bf r}_1)\Delta ({\bf r}_2)
\Delta ({\bf r}_3)> = {\Delta}^{4}
(W_G(|{\bf r}- {\bf r}_{1}|)
W_G(|{\bf r}_2- {\bf r}_3|) + W_G(|{\bf r}- {\bf r}_2|)
W_G(|{\bf r}_1- {\bf r}_3|)
$$
$$
 + W_G(|{\bf r}- {\bf r}_3|)
W_G(|{\bf r}_1- {\bf r}_2|)). \eqno (17.c)
$$

For this particular medium in the sum $T_3$ all the terms $w^{(2n-
1)} = 0$. We use Eq.(11) and Eq.(12) to calculate $w^{(2)}$ and 
$w^{(4)}$. Then 
$$
w^{(2)} = \frac{\Delta ^2}{9}\int {\rm d}^{3}k W_G(k). \eqno (18)
$$
Being the value of $W_G(r)$ when $r=0$, the integral in right-hand 
side of Eq.(18) is equal to one. Thus, $w^{(2)} = {\Delta}^2/9$. 
Note, Eq.(18) is valid not only in the case of Gaussian 
inhomogeneities, but for any random zero-mean function 
$\Delta ({\bf r})$. Consequently, always the first (quadratic in 
fluctuations of LC) correction to EC is independent of the form of 
the correlation function (see, for example, \cite{13}).

Our calculations showed that $w^{(4)} = {\Delta}^{4}(\pi - 
20/9)/9$. Then according to Eq.(14) and Eq.(15) up to the terms of 
the order of ${\Delta}^{4}$ we have 
$$
T_3^{(G)} = 1 + \frac{1}{9}\left\{{\Delta}^{2} + 
{\Delta}^{4}(\pi - \frac{20}{9})\right\}, \quad
{\sigma}_{ef}^{(G)} = <\sigma >\left\{1-\frac{1}{3}[{\Delta}^{2} + 
{\Delta}^{4}(\pi - \frac{7}{3})]\right\}. \eqno (19)
$$

For Gaussianly distributed random function $\Delta ({\bf r})$ it 
is rather simple to calculate ${\sigma}_{ef}^{(G)}$  up to the 
higher order terms in ${\Delta}^{2}$. However, we failed trying 
to obtain a general formula for these terms. 

\section{The Effective Conductivity of Polycrystals}

Polycrystals are widespread case of inhomogeneous media, where the 
inhomogeneity is due to different orientations of discrete single 
crystal grains. Anisotropic properties of each grain are described 
by tensor characteristics. If crystallographic axes of the grains 
are randomly rotated with respect to a fixed set of laboratory 
axes, the characteristics of the medium measured in the laboratory 
coordinate system are stochastic tensor functions of position. 
However, the invariants of the tensors are the same for all the 
grains and do not depend on position. Consequently, effective 
characteristics of polycrystals have to be expressed in terms of 
the invariants of the tensors that describe the phenomenon. When a 
polycrystal is isotropic in average (it is untextured), its 
effective characteristics are isotropic tensors. For example, EC 
of an untextured polycrystal is ${\sigma}_{ik}^{ef} = 
{\sigma}_{ef}{\delta}_{ik}$.

\subsection{Two-dimensional Polycrystals}

For 2-D polycrystal there is the exact solution for 
${\sigma}_{ef}$ found by Dykhne \cite{2}. We use the Dykhne 
formula to verify the calculations of Section 2, comparing the 
perturbation theory results with the expansion of the exact 
solution in powers of the parameter defining the anisotropy of LC. 
We do not find out a procedure allowing us to calculate an 
arbitrary term of the perturbation series. Therefore we only 
checked the first and the second terms taking account of spatial 
fluctuations. This exercise, being a test for our theory, outlined 
the way of calculation of high-order terms in the expression for 
EC of 3-D polycrystals. 

Let ${\sigma}_{ik}$ be the single crystal 2-D conductivity tensor, 
and let ${\sigma}_{1}$ and ${\sigma}_{2}$ be its principle values. 
We set 
$$
\overline \sigma = \frac{1}{2}({\sigma}_{1}+{\sigma}_{2}), \quad
\Delta =  \frac{({\sigma}_{1}-
{\sigma}_{2})}{({\sigma}_{1}+{\sigma}_{2})}, \eqno (20)
$$
where $\overline \sigma $ is the mean conductivity, and the 
dimensionless parameter $\Delta$ ($|\Delta|<1$) defines the 
deviation of the principle conductivities from 
$\overline \sigma$: ${\sigma}_{1} = \overline \sigma(1 + \Delta)$ 
and ${\sigma}_{2} = \overline \sigma (1 - \Delta)$. Being the 
first order invariant of the tensor ${\sigma}_{ik}$, the mean 
conductivity is independent of position. Evidently, $\overline 
\sigma = <\sigma>$. 

According to Dykhne, EC of a 2-D polycrystal does not depend on 
correlations between stochastic functions ${\Delta}_{ik}({\bf r})$ 
at different positions: ${\sigma}_{ef} = 
\sqrt{{\sigma}_{1}{\sigma}_{2}}$. Note, ${\sigma}_{1}{\sigma}_{2}$ 
being the determinant of the tensor ${\sigma}_{ik}$, is its 
invariant. 

We rewrite the Dykhne formula as 
$$
{\sigma}_{ef} = \overline\sigma\sqrt {(1-{\Delta}^{2})}
=\overline\sigma (1 - \frac{{\Delta}^{2}}{2} - 
\frac{{\Delta}^{4}}{8}...) \eqno (21)
$$
and compare the terms of the series (21) with the result obtained 
with the aid of Eqs.(16).

To start the calculation we write the local conductivity tensor 
with respect to the laboratory coordinate system: ${\sigma}_{ik} = 
= \overline\sigma ({\delta}_{ik} + {\Delta}_{ik}({\bf r}))$, where  
$$
{\Delta}_{ik}({\bf r})=\Delta Z_{ik}({\bf r}); \quad 
Z_{ik}=({\alpha}_{i1}{\alpha}_{k1}-
{\alpha}_{i2}{\alpha}_{k2}),  \eqno (22)
$$ 
and ${\alpha}_{ik}= {\alpha}_{ik}({\bf r})$ is 2-D rotation matrix 
that defines the orientation of the crystallographic axes of the 
grain containing the point $\bf r$ with respect to the laboratory 
axes. The elements of 2-D rotation matrix depend on one angle 
only. Let it be the angle $\psi$. The value of $\psi$ is a 
stochastic function of position.

With regard to Eqs.(16) when calculating EC of 2-D polycrystals up 
to the terms of the order of ${\Delta}^4$, we need to calculate 
the terms $v^{(n)}$ ($n = 2,3,4$) of the series (16.a) for 
$T_2$. Consequently, first, we must define the second, the 
third and the fourth moments of the stochastic tensor function 
${\Delta}_{ik}({\bf r})$. The general statistical properties of 
polycrystals were discussed in Ref.\cite{9}. Here we discuss them 
briefly, paying attention to calculation of the high-order 
moments. 

The only property of the medium that affects the ensemble averages 
is the rotations of the crystallographic axes of the grains. Then 
ensemble average becomes the average over all possible rotations 
of the crystallites. We assume that in the ensemble the angles 
defining the rotations of different grains are statistically 
independent. When calculating the second order two-point 
correlator $<{\Delta}_{ik}({\bf r}){\Delta}_{li}({\bf r}_1)>$ 
entering the expression for $v^{(2)}$, there are two cases to 
consider:

1. $\bf r$ and ${\bf r}_{1}$ are in the same grain;

2. $\bf r$ and ${\bf r}_{1}$ are in different grains.

\noindent  We denote the probability of the case 1 by 
$W_2(|{\bf r}-{\bf r}_{1}|)$. Then $1-W_2$ is the probability of 
the case 2. The fact that the probability $W_2$ depends on $\bf r$ 
and ${\bf r}_{1}$ only through $|{\bf r}-{\bf r}_{1}|$ is the 
consequence of our assumption that the polycrystalline medium is 
isotropic and statistically homogeneous. 

Let ${\psi}_n$ be the angle defining the orientation of the 
crystallographic axes in the grain containing the point ${\bf 
r}_n$. In the case 1 the rotations defined by the angles $\psi$ 
and ${\psi}_{1}$ are identically equal. So, these both angles 
define one specific rotation. In this case the second order two-
point correlator $<{\Delta}_{ik}({\bf r}){\Delta}_{li}({\bf 
r}_1)>$ reduces to the second order one-point correlator 
${S}_{ik;li} = <{\Delta}_{ik}({\bf r}){\Delta}_{li}({\bf r})>$. 
This is an isotropic second-rank tensor: 
${S}_{ik;li}=S{\delta}_{kl}$. As far as ${\Delta}_{ii} = 0$, 
${\Delta}_{ik}{\Delta}_{ki}=2{\Delta}^{2}$ is the only nonzero 
second order invariant of the tensor ${\Delta}_{im}({\bf r})$. 
Then $S = {\Delta}^{2}$.

Because all the rotations are assumed to be independent, in the 
case 2 we have $<{\Delta}_{ik}({\bf r}){\Delta}_{li}({\bf r}_1)> 
= <{\Delta}_{ik}({\bf r})><{\Delta}_{li}({\bf r}_1)>=0$. 
Consequently, the case 2 does not contribute to the value of the 
second order two-point correlator  
$<{\Delta}_{ik}({\bf r}){\Delta}_{li}({\bf r}_1)>$. Thus
$$
<{\Delta}_{ik}({\bf r}){\Delta}_{li}({\bf r}_1)> = {\Delta}^{2}
W_2(|{\bf r}-{\bf r}_{1}|){\delta}_{kl}. \eqno (23)
$$

Next, it is evident that the three-point average 
$<{\Delta}_{ik}({\bf r}){\Delta}_{lm}({\bf r}_1){\Delta}_{ni}({\bf 
r}_2)>$ is not equal to zero only if all the three vectors 
$\bf r$, ${\bf r}_{1}$ and ${\bf r}_{2}$ are in the same grain. 
Let us denote the probability of this event as $W_3 = 
W_3(|{\bf r}-{\bf r}_{1}|,|{\bf r}-{\bf r}_{2}|, 
|{\bf r}_{1}-{\bf r}_{2}|)$. Then 
$$
<{\Delta}_{ik}({\bf r}){\Delta}_{lm}({\bf r}_1){\Delta}_{ni}({\bf 
r}_2)> = {S}_{ik;lm;ni} W_3(|{\bf r}-{\bf r}_{1}|,|{\bf r}-
{\bf r}_{2}|,|{\bf r}_{1}-{\bf r}_{2}|), \eqno (24)
$$
where ${S}_{ik;lm;ni}= <{\Delta}_{ik}({\bf r}){\Delta}_{lm}
({\bf r}){\Delta}_{ni}({\bf r})>$ is the third order one-point 
average. It is clear that ${S}_{ik;lm;ni}$ is the isotropic forth-
rank tensor. One can easily verify that in the 2-D case all the 
invariants of the tensor ${\Delta}_{ik}{\Delta}_{lm}{\Delta}_{ni}$ 
are equal to zero. Then, without calculations we can state that 
all the elements of the tensor ${S}_{ik;lm;ni}$ vanish. As a 
result, $v^{(3)}=0$, and the third order correction to EC 
vanishes. 

Finally, it is clear that there are only two possibilities when 
the four-point average $<{\Delta}_{ik}({\bf r}){\Delta}_{lm}
({\bf r}_1){\Delta}_{np}({\bf r}_2){\Delta}_{qi}({\bf r}_3)>$ is 
not equal to zero. Namely,

1.Each of the pairs of the four vectors, but not all of 
them simultaneously, are in the same grain. 

2. All the four vectors are in the same grain.

For the case 1 by $W_4([{\bf r}_a,{\bf r}_b],[{\bf r}_c,{\bf 
r}_d])$ we denote the joint conditional probability for the 
vectors ${\bf r}_a$ and ${\bf r}_b$ to get in the same grain, and, 
simultaneously, for the vectors ${\bf r}_c$ and ${\bf r}_d$ to get 
in some other grain. The probability $W_4([{\bf r}_a,{\bf 
r}_b],[{\bf r}_c,{\bf r}_d])$ excludes the possibility for all the 
four vectors to be in the same grain. Evidently, since all the 
rotations of the grains are independent, in the case 1 the four-
point average is a product of two one-point averages of the second 
order. Next, by $W_4([{\bf r},{\bf r}_{1},{\bf r}_{2},{\bf 
r}_{3}])$ we denote the probability of the case 2. Then 
$$
<{\Delta}_{ik}({\bf r}){\Delta}_{lm}({\bf r}_1){\Delta}_{np}
({\bf r}_2){\Delta}_{qi}({\bf r}_3)> = 
W_4([{\bf r},{\bf r}_{1}],[{\bf r}_{2},
{\bf r}_{3}]){S}_{ik,lm}{S}_{np,qi} + 
W_4([{\bf r}, {\bf r}_{2}],[{\bf r}_{1},
{\bf r}_{3}]){S}_{ik,np}{S}_{lm,qi} 
$$
$$
+ W_4([{\bf r}, {\bf r}_{3}],[{\bf r}_{1},
{\bf r}_{2}]){S}_{ik,qi}{S}_{lm,np} +
W_4([{\bf r}, {\bf r}_{1},{\bf r}_{2},
{\bf r}_{3}]){R}_{ik,lm,np,qi},
\eqno (25.a)
$$
where
$$
{S}_{ik,lm}\;=\;<{\Delta}_{ik}({\bf r}){\Delta}_{lm}({\bf r})>,
\eqno (25.b)
$$
$$
{R}_{ik,lm,np,qi}\; = \; <{\Delta}_{ik}({\bf r})
{\Delta}_{lm}({\bf r})({\bf r})
{\Delta}_{np}({\bf r}){\Delta}_{qi}({\bf r})>. \eqno (25.c)
$$. 

Let us calculate the elements of the isotropic tensors 
${S}_{ik;lm}$ and ${R}_{ik,lm,np,qi}$. Taking account of the 
symmetry properties with respect to interchanges of the indices, 
we see that the isotropic forth rank tensor ${S}_{ik;lm}$ must 
have the form $S_{ik,lm}=a{\delta}_{ik}{\delta}_{lm} + 
b({\delta}_{il}{\delta}_{km} + {\delta}_{im}{\delta}_{kl})$. To 
determine the values of $a$ and $b$ we calculate the two 
independent invariants of the tensor ${\Delta}_{ik}{\Delta}_{lm}$.
They are the contractions ${\Delta}_{kk}{\Delta}_{ll} = 0$ and 
${\Delta}_{ik}{\Delta}_{ki} = 2{\Delta}^{2}$. Then 
$$
{S}_{ik;lm} = \frac{{\Delta}^{2}}{2}(-{\delta}_{ik}{\delta}_{lm}+
{\delta}_{il}{\delta}_{km}+{\delta}_{im}{\delta}_{kl}).  
\eqno (26.a)
$$
If we use the explicit form of the tensor ${\Delta}_{ik}$ 
(see Eq.(22)), it is easy to see that in the 2-D case the sixth 
rank tensor
$$
{R}_{ik,lm,np,qi}\; = \; {S}_{ik,qi}{S}_{lm,np}\; = 
\;{\Delta}^{2}{\delta}_{kq}{S}_{lm,np}. 
\eqno (26.b)
$$

Now we are ready to calculate the terms  $v^{(2)}$ and $v^{(4)}$ 
of the sum $T_2$. The same as when deriving Eq.(18), with the aid 
of Eq.(23) we obtain $v^{(2)} = {\Delta}^{2}/4$. As usual, the 
first correction term is independent of the form of the 
correlation function $W_2(r)$. 

To calculate $v^{(4)}$ we substitute Eqs.(25) and Eqs.(26) in 
Eq.(16.b) for $n=4$ and calculate the contraction entering the 
integrand of the left-hand side of this equation. Then 
$$
v^{(4)} = \frac{{\Delta}^{4}}{16}\sum_{q=1}^{3}X^{(q)}.
\eqno (27.a)
$$
where
$$
X^{(1)} = \int\int \int {\rm d}^{2}r_{1}{\rm d}^{2}r_{2}{\rm 
d}^{2}r_{3}
\delta ({\bf r}-{\bf r}_{1})\delta ({\bf r}_1-{\bf r}_{2}) 
\delta ({\bf r}_2-{\bf r}_{3})
W_4([{\bf r},{\bf r}_{1}],[{\bf r}_{2},{\bf r}_{3}]), \eqno (27.b)
$$
$$
X^{(2)} = \frac{1}{{(2\pi)}^{4}}\int\int \int {\rm d}^{2}r_{1}
{\rm d}^{2}r_{2}
{\rm d}^{2}r_{3}\int \int{\rm d}^{2}k_{1}{\rm d}^{2}k_{3}
{\rm e}^{i{\bf k}_1({\bf r}-{\bf r}_1)}
{\rm e}^{i{\bf k}_3({\bf r}_2-{\bf r}_{3})} \times
$$
$$
[2{({\vec \kappa}^{(1)}{\vec \kappa}^{(3)})}^{2} - 1]
\delta ({\bf r}_1-{\bf r}_{2}) 
W_4([{\bf r}, {\bf r}_{2}],[{\bf r}_{1},{\bf r}_{3}]), 
\eqno (27.c)
$$
and
$$
X^{(3)} = \frac{2}{{(2\pi)}^{4}}\int\int \int {\rm d}^{2}r_{1}{\rm 
d}^{2}r_{2}
{\rm d}^{2}r_{3}\int \int{\rm d}^{2}k_{1}{\rm d}^{2}k_{3}
{\rm e}^{i{\bf k}_1({\bf r}-{\bf r}_1)}
{\rm e}^{i{\bf k}_3({\bf r}_2-{\bf r}_{3})} \times
$$
$$
{({\vec \kappa}^{(1)}{\vec \kappa}^{(3)})}^{2}\delta ({\bf r}_1-
{\bf r}_{2}) 
[W_4([{\bf r}, {\bf r}_{3}],[{\bf r}_{1},{\bf r}_{2}]) +
W_4([{\bf r}, {\bf r}_{1},{\bf r}_{2},{\bf r}_{3}])] . 
\eqno (27.d)
$$

Because of the presence of the $\delta$-functions in the integrand 
of the expression for $X^{(1)}$, we have $X^{(1)} = W_4([{\bf 
r},{\bf r}],[{\bf r},{\bf r}])$. Thus, ${\bf r} = {\bf r}_{1} = 
{\bf r}_{2} = {\bf r}_{3}$, and, consequently, all the four 
vectors are in the same grain. However, according to the 
definition, the probability $W_4([{\bf r}_a,{\bf r}_b],
[{\bf r}_c,{\bf r}_d])$ is not equal to zero only if the vectors 
${\bf r}_a$ and ${\bf r}_b$ are in the same grain, but the vectors 
${\bf r}_c$ and ${\bf r}_d$ are in some other grain. Then $X^{(1)} 
= 0$. The same argumentation is valid when calculating $X^{(2)}$. 
Really, after integration over ${\bf r}_{2}$, the probability 
$W_4([{\bf r}, {\bf r}_{1}],[{\bf r}_{1},{\bf r}_{3}])$ enters the 
integrand. This probability is equal to zero too. Then $X^{(2)} = 
0$.

As a result, only the third term $X^{(3)}$ of the sum (27.a) 
contributes to $v^{(4)}$. In the integrand of Eq.(27.d) the sum of 
probabilities $W_4([{\bf r},{\bf r}_{3}],[{\bf r}_{1},
{\bf r}_{2}]) + W_4([{\bf r}, {\bf r}_{1},{\bf r}_{2},{\bf 
r}_{3}])$ defines the combined probability for the pairs 
of the vectors ${\bf r}$, ${\bf r}_{3}$ and ${\bf r}_{1}$, ${\bf 
r}_{2}$ be in the same grains not making difference between the 
cases when the first and the second pairs get in one or distinct 
grains. This combined probability can be expressed in terms of the 
probability $W_2$. Namely, $W_4([{\bf r}, {\bf r}_{3}],[{\bf 
r}_{1},{\bf r}_{2}]) + W_4([{\bf r}, {\bf r}_{1},{\bf r}_{2},{\bf 
r}_{3}]) = W_2(|{\bf r} - {\bf r}_{3}|)W_2(|{\bf r}_{1}-{\bf 
r}_{2}|)$. When substituting this equality in Eq.(27.d) and 
performing the integrations, we obtain $v^{(4)} = {\Delta}^{4}/8$.

Finally, with regard to Eqs.(18) up to the terms of the order of 
${\Delta}^{4}$ we have
$$
T_2 = 1 + \frac{{\Delta}^{2}}{4} + \frac{{\Delta}^{4}}{8} \quad 
{\rm and} \quad {\sigma}_{ef} = \overline\sigma (1 - 
\frac{{\Delta}^{2}}{2} - \frac{{\Delta}^{4}}{8}). \eqno (28)
$$
(compare with Eq.(21)). We see that within the accuracy of 
calculation EC is the same as the one defined by the Dykhne 
formula.

The results of this subsection can be formulated as follows: we 
verified the perturbation theory of Section 2 and find out the 
forms of the correlators (see Eqs.(23), (24) and (25)) entering 
the integrands of the terms of the series $T_2$. With regard to 
the dimensionality of the problem, the same form of the 
correlators must be used when calculating EC of 3-D polycrystals. 

\subsection{Three-dimensional Polycrystals}

Let a polycrystal be composed from 3-D single crystal grains. The 
local conductivity tensor is defined by Eq.(1). If the polycrystal 
is isotropic in average, $<\sigma> = \overline\sigma = 
({\sigma}_{1} + {\sigma}_{2}+{\sigma}_{3})/3$, ${\sigma}_{i}$ 
($i=1,2,3$) are the principal values of the single crystal 
conductivity tensor. The principle values of the tensor 
${\Delta}_{ik}$ are 
$$
{\Delta}_{1} = \frac{(2{\sigma}_{1}-{\sigma}_{2}-{\sigma}_{3})}
{({\sigma}_{1}+{\sigma}_{2}+{\sigma}_{3})}; \quad
{\Delta}_{2} = \frac{(2{\sigma}_{2}-{\sigma}_{1}-{\sigma}_{3})}
{({\sigma}_{1}+{\sigma}_{2}+{\sigma}_{3})}; \quad
{\Delta}_{3} = \frac{(2{\sigma}_{3}-{\sigma}_{1}-{\sigma}_{2})}
{({\sigma}_{1}+{\sigma}_{2}+{\sigma}_{3})}.  \eqno (29.a)
$$
Since ${\Delta}_{1}+{\Delta}_{2}+{\Delta}_{3}=0$, the anisotropy 
of LC of a 3-D polycrystal is defined by two independent 
parameters. With respect to the laboratory coordinate system 
$$
{\Delta}_{ik}={\Delta}_{1}{\alpha}_{i1}{\alpha}_{k1} +
{\Delta}_{2}{\alpha}_{i2}{\alpha}_{k2} + 
{\Delta}_{3}{\alpha}_{i3}{\alpha}_{k3},
\eqno (29.b)
$$
where ${\alpha}_{ik}$ is the 3-D rotation matrix, whose elements 
are stochastic functions of position. 

In what follows we calculate the third order fluctuation 
correction to EC of 3-D polycrystals. In contrast to the 2-D case, 
this correction does not vanish. To perform the calculation up we 
need to define the terms $w^{(2)}$ and $w^{(3)}$ of the series 
$T_3$ (see Eq.(14) and Eq.(15)).

We calculate the two-point average entering the expression for 
$w^{(2)}$ the same as in Subsection 3.1. With regard to the 
dimensionality of the problem 
$<{\Delta}_{ik}({\bf r}){\Delta}_{li}({\bf r}_1)> = 
D_2 W_2(|{\bf r}-{\bf r}_{1}|){\delta}_{kl}/3$ where $D_2 $ is the 
second-order invariant of the tensor ${\Delta}_{ik}$: ${D}_{2} = 
{\Delta}_{ik}{\Delta}_{ki}$. Then with the aid of Eq.(11) and 
Eq.(12) we have $w^{(2)} = D_2/27$. Again the quadratic term of 
the series $T_3$ and, consequently, the quadratic correction to EC 
do not depend on the form of the correlation function.

When calculating $w^{(3)}$ we use Eq.(24) to calculate the three-
point average $<{\Delta}_{ik}({\bf r}){\Delta}_{lm}({\bf r}_1)
{\Delta}_{ni}({\bf r}_2)>$. This average does not vanish only if 
all the three vectors  $\bf r$, ${\bf r}_{1}$ and ${\bf r}_{2}$ 
are in the same grain. As in the 2-D case, the elements of the 
isotropic forth rank tensor ${S}_{ik;lm;ni}$ have to be expressed 
in terms of the invariants of the tensor 
${\Delta}_{ik}{\Delta}_{lm}{\Delta}_{ni}$. However, 
in contrast to the 2-D case, only one of two independent 
invariants of this tensor is equal to zero. Namely, 
${\Delta}_{ik}{\Delta}_{ki}{\Delta}_{ll} = 0$, but 
${\Delta}_{ik}{\Delta}_{kl}{\Delta}_{li} = D_3$, 
where $D_3 ={\Delta}_{1}^3 + {\Delta}_{2}^3 + {\Delta}_{3}^3$. 
Then taking into account the symmetry properties of the tensor 
${S}_{ik;lm;ni}$ we have
$$
{S}_{ik;lm;ni} = \frac{D_3}{30}[-2{\delta}_{kn}{\delta}_{lm} + 
3({\delta}_{kl}{\delta}_{nm} + {\delta}_{km}{\delta}_{nl})]. 
\eqno (30.a)
$$

The probability $W_3$ for the three vectors $\bf r$, ${\bf r}_{1}$ 
and ${\bf r}_{2}$ to be in the same grain can be expressed in 
terms of the probabilities $W_2$. Indeed, let us choose one of the 
three vectors ${\bf r}$, ${\bf r}_1$ and ${\bf r}_2$. Since all 
the choices are equiprobable, the probability to choose one 
definite vector is equal to $1/3$. Let it be the vector $\bf r$. 
Then the joint conditional probability of choosing the vector 
$\bf r$, and finding vector ${\bf r}_1$ in the same grain 
with the vector $\bf r$, and finding vector ${\bf r}_2$ in the 
same grain with the vector $\bf r$ is 
$W_2(|{\bf r}-{\bf r}_{1}|)W_2(|{\bf r}-{\bf r}_{2}|)/3$. Now it 
is easy to see that 
$$
W_3 = \frac{1}{3}( W_2(|{\bf r}-
{\bf r}_{1}|)W_2(|{\bf r}-{\bf r}_{2}|) +
W_2(|{\bf r}_{1}-{\bf r}|) W_2(|{\bf r}_{1}-{\bf r}_{2}|) + 
W_2(|{\bf r}_{2}-{\bf r}|)W_2(|{\bf r}_{2}-{\bf r}_{1}|)). 
\eqno (30.b)
$$

When we take into account Eqs.(30) and substitute Eq.(24) in 
Eq.(12), after integration over position vectors and over the 
angles in the double integral over the wave vectors ${\bf k}_{1}$ 
and ${\bf k}_{2}$, we obtain
$$
w^{(3)} = -\frac{D_3}{405}\left\{ \frac{17}{3} - J\right\},
\eqno (31.a)
$$
$$
J = 2{\pi}^{2}\int_{0}^{\infty}\int_{0}^{\infty}{\rm d}k_1{\rm 
d}k_2 W_2(k_1)W_2(k_2)\left\{ 1- \frac{{(k_1^2-
k_2^2)}^{2}}{2k_1q_2(k_1^2+
k_2^2)}\ln |(k_1+k_2)/((k_1-k_2))|\right\}.  \eqno (31.b)
$$
Then up to the third order term in the anisotropy we have
$$
T_3 = 1 + w^{(2)} + w^{(3)}, \quad
 {\sigma}_{ef} = \overline\sigma \left[ 1 - \frac{D_2}{9} 
+\frac{D_3}{135}\left(\frac{17}{3} - J\right)\right].  \eqno (32)
$$

From Eq.(32) it is clearly seen that EC of a 3-D polycrystal 
depends on the statistical properties of the medium. Of course, 
this is the well-known result. However, now we can estimate the 
influence of the form of the correlation function on the value of 
${\sigma}_{ef}$. To do this we evaluated $J$ for 
several choices of $W_2(r)$. The obtained results were

1. $J = 0.028$, when $W_2 = {\rm exp}(-r^2/a^2)$;

2. $J = 0.136$, when $W_2 = {\rm exp}(-r/a)$;

3. $J = 0.052$, when $W_2 = 1/(1+r^2/a^2)$.

\noindent For all the examined forms of the correlation function the value 
of $J$ is very small. Consequently, although the value of the third order term
in the  expression for EC depends on the form of the correlation function, 
this dependence is very weak. 

Concluding this subsection, let us estimate when the low order 
perturbation theory calculations are sufficient. In Ref.\cite{14} 
it was shown that the effective conductivity is restricted by 
inequalities $<{\bf j}^2>/{\bf j}{\hat\sigma}^{-1}{\bf j} \le 
{\sigma}_{ef} \le {\bf e}\hat\sigma{\bf e}/<{\bf e}^{2}>$, where 
${\hat\sigma}^{-1}$ is the inverse of the local conductivity 
tensor $\hat\sigma$; ${\bf j}$ and ${\bf e}$ are arbitrary test 
vectors of dimension of the current density and the electric 
field, respectively. If as the test vectors we choose a direct 
current ${\bf j}_0$ and a uniform electric field ${\bf e}_0$, we 
obtain
$$
\frac{1}{<\rho>} \; \le \;{\sigma}_{ef}\; \le\; <\sigma>. 
\eqno (33)
$$
In polycrystals $<\sigma>$ is equal to the mean conductivity 
$({\sigma}_{1} + {\sigma}_{2} + {\sigma}_{3})/3$ and $<\rho>$ is 
equal to the mean resistivity $({\sigma}_{1}^{-1} + 
{\sigma}_{2}^{-1} + {\sigma}_{3}^{-1})/3$.

For simplicity, let us examine a uniaxial metal: ${\sigma}_{1} = 
{\sigma}_{2}$. According to Eq.(29.a) in this case 
${\Delta}_{1}={\Delta}_{2}=\Delta = ({\sigma}_{1}-
{\sigma}_{3})/(2{\sigma}_{1}+{\sigma}_{3})$, and ${\Delta}_{3}=-
2\Delta$. Then $-1 < \Delta < 1/2$, when $0 <  
{\sigma}_{1}/{\sigma}_{3} < \infty$. In Fig.1 following Eq.(33) we 
plot the lower and the upper bound of the ratio 
${\sigma}_{ef}/\overline\sigma$ as function of $\Delta$ as well as 
the values of ${\sigma}_{ef}/\overline\sigma$ calculated with the 
aid of Eq.(32). If $-.2 < \Delta < .2$ (or $4/7 < 
{\sigma}_{1}/{\sigma}_{3} < 2$), the difference 
between the lower bound for EC and its upper bound is less than 10 
percents. It is reasonable to assume that the perturbation theory 
results are applicable not only if $|{\sigma}_{1}/{\sigma}_{3}-1| 
\ll 1$, but at least for all ${\sigma}_{1}/{\sigma}_{3}$ from the 
aforementioned interval. The Table 1 based on date from \cite{15} 
shows that rather often the perturbation theory calculations are 
sufficient when estimating EC. 

We would like to note that when comparing the calculated value of 
${\sigma}_{ef }$ with experimental results, one must have in mind 
that in our calculations only the inhomogeneity due to different 
orientations of polycrystalline grains was taken into account. Of 
course, in real polycrystals there are other sources of 
inhomogeneity too. For example, we do not take into account the 
real structure of the boundaries of the grains. This 
simplification is justified when the grains are sufficiently large 
and the properties of the boundaries of the grains do not affect 
the result significantly.

\centerline{\bf ACKNOWLEDGMENTS} 

The author is grateful to Prof. A.M.Dykhne and to Prof. 
M.I.Kaganov for helpful comments during the course of this work.
The work was supported by RBRF Grant No. 02-02-17226.

\newpage
\centerline{\bf List of Figures}

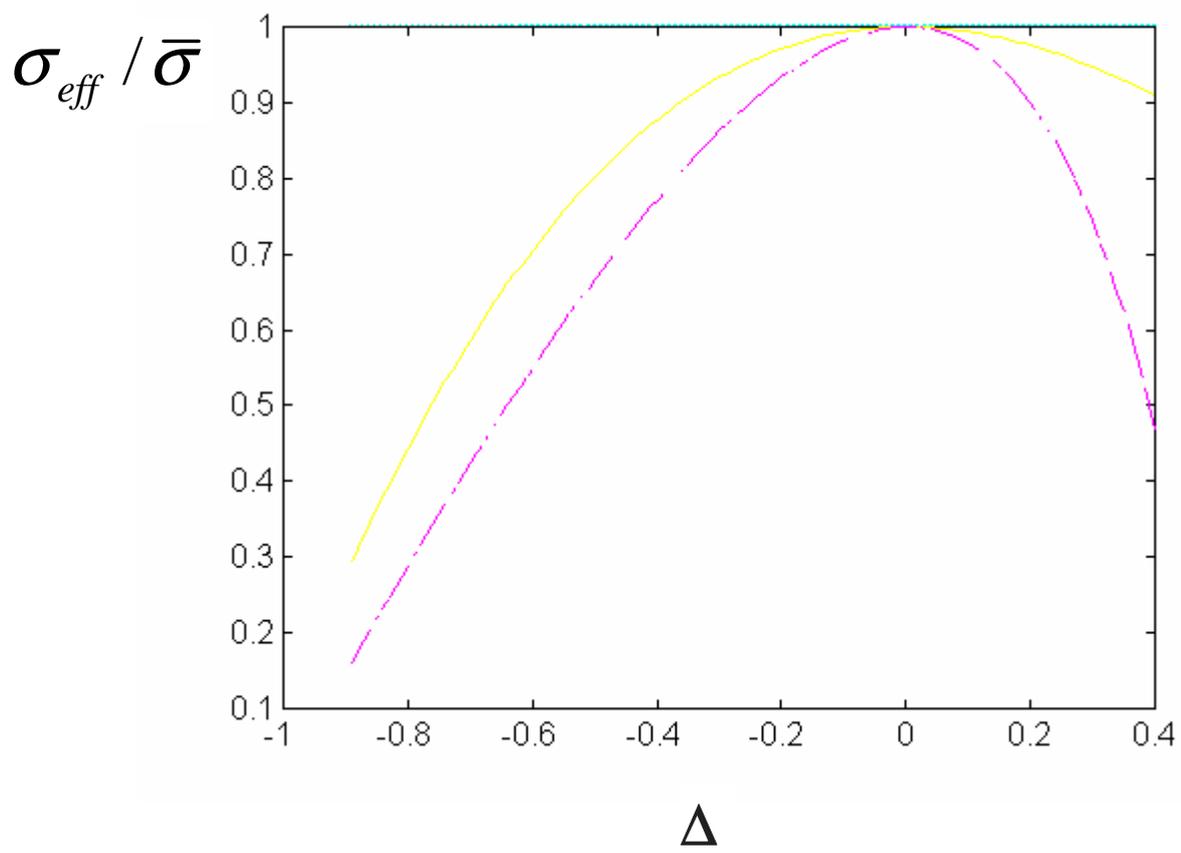
Fig.1. The lower bound (dashed line) and the value of 
${\sigma}_{ef}/\overline\sigma$ (full line) for uniaxial 
polycrystals as functions of the parameter $\Delta$. Always the 
ratio ${\sigma}_{ef}/\overline\sigma < 1$. The value of 
${\sigma}_{ef}/\overline\sigma$ is calculated with the accuracy
up to ${\Delta}^3$.

\end{document}